\documentclass[runningheads,dvipsnames]{llncs}

\usepackage[utf8]{inputenc} \usepackage[T1]{fontenc}
\usepackage{graphicx}

\usepackage{hyperref} \usepackage{tcolorbox} \usepackage{xcolor}

\usepackage{tikz} 
\usetikzlibrary{positioning, fit, calc}
\usetikzlibrary{backgrounds}
\usepackage[backend = biber, style = lncs, natbib = true, hyperref = true, url = true%,
]{biblatex}

\DeclareNameFormat{author}{%
  \ifdefvoid{\namepartprefix}{}{\namepartprefix\space}\namepartfamily,
  \namepartgiveni%
  \ifthenelse{\value{listcount}<\value{liststop}} {\addcomma\space}%
  {}%
} \DeclareNameFormat{editor}{%
  \ifdefvoid{\namepartprefix}{}{\namepartprefix\space}\namepartfamily,
  \namepartgiveni%
  \ifthenelse{\value{listcount}<\value{liststop}} {\addcomma\space}%
  {\space\ifthenelse{\value{listcount}>1} {(\bibstring{editors})}
    {(\bibstring{editor})}}%
} 
\renewbibmacro{finentry}{\settoggle{lncs:lncs}{false}\finentry}
\DeclareBibliographyAlias{inbook}{inproceedings}
\DeclareFieldFormat[inbook]{title}{#1}
\newcommand{\noshow}[1]{}

\newcommand{\acronym}{\textsc{Reasonable Machines\/}}

\newcommand{\MI}{Responsible Machine Architecture}
\newcommand{\MII}{Ethico-Legal Ontologies} \newcommand{\MIII}{Symbolic
  Reasoning Tools} \newcommand{\MIV}{Interpretable AI Systems}
\newcommand{\MV}{Human-Machine Communication \& Interaction}
\newcommand{\MVI}{Cloud-based Reasoning Workbench}
\newcommand{\MVII}{Use Cases and Empirical Studies}

\newcommand{\logikey}{\textsc{LogiKEy}}

\begin{document}

\title{Reasonable Machines: A Research Manifesto}

\author{Christoph Benzmüller\inst{1}\orcidID{0000-0002-3392-3093} \and
  Bertram Lomfeld\inst{2}\orcidID{0000-0002-4163-8364}
  % \and David Fuenmayor\inst{1}\orcidID{0000-0002-0042-4538}
}
\authorrunning{C.~Benzmüller and B.~Lomfeld}

\institute{Institute of Computer Science, Freie Universit\"at Berlin,
  Berlin, Germany %\email{c.benzmueller@fu-berlin.de} 
  \and Department
  of Law,
  Freie Universit\"at Berlin, Berlin, Germany\\
  \email{c.benzmueller|bertram.lomfeld@fu-berlin.de}}

\maketitle % typeset the header of the contribution

\begin{abstract}
  Future intelligent autonomous systems (IAS) are inevitably deciding
  on moral and legal questions, e.g. in self-driving cars, health care
  or human-machine collaboration. As decision processes in most modern
  sub-symbolic IAS are hidden, the simple political plea for
  transparency, accountability and governance falls short. A sound
  ecosystem of trust requires ways for IAS to autonomously justify
  their actions, that is, to learn giving and taking reasons for their
  decisions. Building on social reasoning models in moral psychology
  and legal philosophy such an idea of
  \guillemotright\acronym\guillemotleft\ requires novel, hybrid
  reasoning tools, ethico-legal ontologies and associated
  argumentation technology. Enabling machines to normative
  communication creates trust and opens new dimensions of AI
  application and human-machine interaction.

   \end{abstract}

\section{Introduction}
Intelligent autonomous systems (IASs) are rapidly entering
applications in industry, military, finance, governance,
administration, healthcare, etc., leading to a historical transition
period with unprecedented dynamics of innovation and change, and with
unpredictable
outcomes.
Politics, regulatory bodies, indeed society as
a whole, are challenged not only with keeping pace with these
potentially disruptive developments, but also with staying ahead and
wisely guiding the transition. Fostering positive impacts, while
preventing negative side effects, is a balanced vision shared within most of the numerous ethical guidelines of the last years on trustworthy AI, including the European Commission's most
recent White Paper on AI \parencite{WhitePaperEU}, proposing the
creation of an \textit{``ecosystem of excellence''} in combination
with an \textit{``ecosystem of trust''}.

We think that real \textit{``Trustworthy AI by Design''} demands IASs,
which are able to give and take reasons for their decisions to act.
Such \guillemotright\acronym\guillemotleft\ require novel, hybrid
reasoning tools, upper ethico-legal ontologies and associated
argumentation technology to be utilised in practice for assessing,
justifying and controlling (externally and internally) the behaviour
of IASs with respect to explicitly encoded legal and ethical
regulation. We envision this technology to be integrated with an
on-demand, cloud-based workbench for pluralistic, expressive
regulatory reasoning. This would foster knowledge transfer with
industry, research, and educational institutions, it would enable
{access to critical AI infrastructure {at scale} with little
  risk and minimal costs}, and, in the long run, it could support dynamic 
adjustments of regulating code for IASs in the cloud via
politically and socially 
legitimated processes.

Paper structure: Section~\ref{sec:objectives} formulates objectives
for \acronym, and section~\ref{sec:evidence} provides models for them
building on moral psychology and legal philosophy.
Section~\ref{sec:steps} outlines modular steps for research and
implementation of \acronym; this leverages own prior work such as the
\logikey\ methodology and framework for designing normative theories
for ethical and legal reasoning \parencite{J48}, which needs to be
combined and extended with an upper-level value ontology
\parencite{lomfeld19:_gramm_recht} and further domain-level regulatory
theories for the assessment and explanation of ethical and legal
conflicts and decisions in IASs.

\section{\acronym: Objectives} \label{sec:objectives} The need for
some form of “moral machines” \cite{wallach08:_moral} is no science
fiction scenario at all. With the rise of autonomous systems in all
fields of life including highly complex and ethically critical
applications like self-driving cars, weapon systems, healthcare
assistance in triage and pandemic plans, predictive policing, legal
judgement supports or credit scoring tools, involved AI systems are
inevitably confronted with, and deciding on, moral and legal questions.
One core problem with ethical and legal accountability or even
governance of autonomous systems is the hidden decision process (black
box) in modern (sub-symbolic) AI technologies, which hinders
transparency as well as direct intervention. The simple plea for
transparency disregards technological realities or even restrains much needed further developments.\footnote{While interpreting, modeling and explaining the inner functioning of black box AI systems is relevant also with respect to our \acronym\ vision, such research alone cannot completely solve the trust and control challenge. Sub-symbolic AI black box systems
  (e.g.~neural architectures) are suffering from various
  issues (including adversarial attacks and influence of bias in data) which
  cannot be easily eliminated by interpreting, modeling and explaining
  them. Offline, forensic processes are then required such that the
  whole enterprise of turning black box AI systems into fully
  trustworthy AI systems becomes a challenging multi-step engineering
  process, and such an approach is significantly further complicated
  when online learning capabilities are additionally foreseen.}

Inspired by moral psychology and cognitive science, we envision the
solution in the development of independent, symbolic logic based
safety-harnesses in future AI systems \parencite{GreeneRTVW16}. Such
“ethico-legal governors” encapsulate and interact with black box AI
systems, and they will use symbolic AI techniques in order to search
for possible justifications, i.e. reasons, for their decisions and
(intended) actions with regard to some formally encoded ethico-legal
theories defined by regulating bodies. The symbolic justifications
computed at this abstract level thus provide a basis for generating
explanations about why a decision/action (proposed by an AI black box
system) is ethico-legally legitimate and compliant with respect to the
encoded ethico-legal regulation.

Such an approach is complementary to, and as an additional measure
more promising than, explaining the inner (mis-)functioning of the
black box AI system itself. Symbolic justifications in turn enable the
development of further means towards a meaningful and robust control
and towards human-understand\-able explanation and human-machine interaction.
The \acronym\ idea outlines a genuine approach of trustworthiness by
  design proposing, in psychological
terminology \cite{kahnemann13}, a slow, rational (i.e.~symbolic)
{``System 2''} layer in responsible IASs to justify and control
their fast, \textit{``intuitive''}, but opaque (sub-symbolic),
{``System 1''} layer computations.

\acronym\ research aims at analyzing and constructing ways how
intelligent machines could socially justify their actions at abstract
level, i.e. give and take moral and legal reasons for their decisions
to act. Reason is based on reasons. This is true as much for
artificial as for human intelligent agents. The ``practical
reasonableness'' of intelligent agents depends on their moral abilities
to communicate socially acceptable reasons for their behavior
\parencite{habermas81:_theor_handel}. Thus, the 
%search for
exploration of methods and tools enabling machines to generate normative reasons (which may be independent of underlying black box architectures and
opaque algorithms) smoothes the way for more comprehensive artificial
moral agency and new dimensions of human-machine communication.

The {core objectives} of \acronym\ technology are: 
\begin{itemize}
\item enabling argument-based explanations \& justifications of IAS
  decisions, 
\item enabling ethico-legal reasoning about, and public
  critique of, IAS decisions, % \& actions,
\item facilitating political and legal governance of IAS decision
  making, 
\item evolving ethico-legal agency and communicative
  capacity of IASs, 
\item enabling trustworthy human-interaction by
  normative communication, % with IASs,
\item fostering development of novel neuro-symbolic AI architectures.
\end{itemize}

\section{Artificial Social Reasoning Model (aSRM)} \label{sec:evidence} The
black box governance problem has an interesting parallel in human
decision making. Most actual models in moral psychology consider
emotional intuition to be \textit{the} (or at least one) initial driving force
of human action which is only afterwards (or with a second
significantly slower system) rationalized with
reasons~\parencite{haidt01, kahnemann13}.
Within a social framework of giving and taking reasons (e.g.~moral
convention or a legal system) the initial motivation of a single human
agent could be ignored if his actions and his post-hoc reasoning
comply with given social (moral or legal) standards
\parencite{lomfeld17:_emotio_iuris}. Communicating reasons within such a post-hoc
\textit{``Social Reasoning Model'' (SRM)} is not superfluous, but essential, as only they guarantee the coherence of a moral or legal
order in an increasingly pluralistic 
world. The remaining difference is the
relative independence of rational reasoning from the motivational
impulse to act. Even so, in the long run the inner-subjective or
social feedback loop with rational reasons might also change the
agents' motivational (emotional) disposition.

This post-hoc SRM is transferable to AI decision processes as
\textit{``artificial Social Reasoning Model'' (aSRM)}. The black box
of an opaque AI system functions like an AI intuition. Following the
SRM model, transparency is not needed as long as the system generates
post-hoc reasons for its action. Moral and legal accountability and
governance could instead be enabled through symbolic or sub-symbolic aSRMs.

A symbolic solution would try to reconstruct (or justify with an
alternative argument) the intuitive decision of the black box with
deontic logical reasoning applying moral or legal standards. A
pluralistic, expressive ``normative reasoning infrastructure'', such as LogiKEy \parencite{J48}, should e.g.~be able to support this
process.

A sub-symbolic solution could create an independent (second) neural
network to produce reasons for the output of the (first) decision
network (e.g.~autonomous driving control). 
Of course, the structure of this ``reasoning net'' process is again hidden. Yet, if the outcoming
reasons coherently comply with prescribed social and ethico-legal
standards the lack of transparency in the second black box constitutes less of a problem. 

Robust solutions for aSRMs could even seek to %tightly 
integrate and align these two options. Moreover, in both scenarios the introduced feedback loop of 
giving and taking reasons could be integrated as learning environment (self-supervised learning) for the initial, intuitive layer of autonomous decision making, with the eventual effect that differences at both layers 
may gradually dissolve. % \marginpar{\mbox{Mention GANs here?}}

Allowing various kinds of reasons, SRMs \& aSRMs 
advance normative pluralism and may integrate different (machine-)ethical traditions: deontological, consequentialist and virtue ethics. 
``Reasonable pluralism'' in recent moral and political
philosophy defines reasonableness by meta-level procedures like
``reflective equilibrium'' and ``overlapping consensus''
\parencite{rawls01:_justic_fairn} or ``rational discourse''
\parencite{habermas81:_theor_handel}. Contemporary legal philosophy
and theory has enfolded how law could act as democratic real-world
implementation of these meta-procedures, structuring public deliberation and
argumentation over conflicting reasons
\parencite{alexy78:_theor_argum,lomfeld15:_gruend_vertr}. Constructing
a pluralist aSRM substantially widens the mostly consequentialist
contemporary approaches \parencite{bonnefon16, GreeneRTVW16} to machine
ethics and moral IAS.

\section{\acronym: Implementation} \label{sec:steps} The
implementation of \acronym\ requires expertise from different areas:
pluralistic normative reasoning, formal ethics and legal theory,
expressive ontologies and semantic web taxonomies, human-computer
interaction, rule-based systems, automated theorem proving,
argumentation technology, neural architectures and machine learning.
Acknowledging the complexity of each field, \acronym\ research should
complement top-down construction of responsible machine architecture
with bottom-up developments starting from existing works in different
domains. More concretely, we propose a modular and stepwise implementation of our research scheme based on the following modules:

\textbf{M1:} \textbf{\MI}. The vision of an aSRM and its parallel to
human SRM needs to be further explored to
guide and refine the overall architectural design of \acronym\ based
on respective system components responsible for generating
justifications, for conducting compliance checks and for governing
the action executions of an IAS.

\textbf{M2:} \textbf{\MII}. Ethico-legal ontologies
constitute a core ingredient to enable the computation, assessment and
communication of aSRM-based rational justifications in the envisioned
ethico-legal governance components for IASs, and they are also key for
black box independent user-explanations in form of rational arguments.
We propose the development of expressive ethico-legal upper-level ontologies  to guide and connect the encoding of concrete ethico-legal domain-level theories (regulatory codes)~\parencite{DBLP:conf/icail/HoekstraBBB09,C76}. Moreover, we propose the concrete regulatory codes to be complemented with an abstract ethico-legal value ontology, for example, as ``discoursive grammar'' of justification \parencite{lomfeld19:_gramm_recht}.

\textbf{M3:} \textbf{\MIII}. For the implementation of
pluralistic, expressive and paradox-free normative reasoning at the upper-level, the LogiKEy framework \parencite{J48} can e.g.~be adapted and further advanced. {LogiKEy works with shallow semantical embeddings (SSEs) of (combinations of) non-classical logics in classical higher-order logic (HOL). HOL thereby serves as a meta-logic, rich enough to support the encoding of a plurality of ``object logics'' (e.g.~conditional, deontic or  epistemic logics and combinations thereof).
% and plural and adaptable value systems.
The embedded ``object logics'' are used for the iterative, experimental encoding of normative theories.} 
This generic approach shall ideally be integrated with specialized solutions based e.g.~on semantic web reasoning, logic programming, answer set programming, and with formalized argumentation for ethical \parencite{DBLP:journals/ail/Verheij16} or legal  \parencite{R80} systems design.

\textbf{M4:} \textbf{\MIV}. Sub-symbolic solutions to SRM-based
accountability and governance challenge could develop a hidden
reasoning net, which might be trained with legal and ethical
use-cases. Moreover, techniques in ``explainable
AI''~\parencite{10.1145/3236009}
have to be assessed and, if possible, integrated with the symbolic
aSRM tools to be developed in M3 in order to provide guidance to their
computations and search processes. The more information can be
obtained about the particular information bits that trigger the
decisions of the black box systems we want to govern, the easier the
corresponding reasoning tasks, i.e. the search for justifications,
should become in the associated, symbolic aSRM tool.

\textbf{M5:} \textbf{\MV}. The intended aSRM-based justifications generated by the tools developed in M3 and M4 require arguments and rational explanation which are understandable for different AI ecosystems \parencite{rahwan19:_machin}, including human users, collect decision scenarios between machines and independent verification tools. Here, the
development of respective techniques could build on argumentation
theory in combination with recent advances towards a computational hermeneutics \parencite{B19}. An overarching objective of \acronym\ is to contribute to trustful and fruitful interaction between human and IASs.

\textbf{M6:} \textbf{\MVI}. To facilitate access to the proposed knowledge representation and reasoning solutions, and also to host the ethico-legal theories, a cloud-based reasoning workbench should be implemented. This workbench
would (i) integrate the bottom-up construed components and tools from
M2-M5 and (ii) implement instances of the top-down governance
architecture(s) developed in M1 based on (i).
This cloud-based solution could be developed in combination with, or as
an alternative to, more independent solutions based e.g. on agent-based development frameworks \parencite{C45}.

\textbf{M7:} \textbf{\MVII}. The overall system framework needs to be
adequately prepared to support changing use cases and empirical
studies. Concrete use cases with high ethical and legal potential
must be defined and employed to guide the research and development
work, as for example the representative issue on
self-driving cars \parencite{bonnefon16}. Empirical studies should support and inform the constructive development process. For testing the ethico-legal 
value ontology in M2, for example, we could try to demonstrate that it can make sense out of the rich  MIT Moral Machine experiment data \parencite{awad18:_moral_machin}. When its architecture evolves, it would be highly valuable to design a genuine aSRM experiment.

\section{Conclusion}
The \acronym\ vision and research requires the integration of heterogeneous and interdisciplinary expertise to be fruitfully
implemented. The cloud-based framework we envision would ideally be
widely available and reusable, and it could become part of related, bigger initiatives towards the sharing of critical AI infrastructure (such as 
the \url{claire-ai.org} vision towards a CERN for AI).
The implementation of the depicted program requires substantial
resources and investment in foundational AI research and in practical system development, but it reflects the urgent and timely need for
the development of trustworthy AI technology.

The possible outreach of the \acronym\ idea is even far
beyond an ecosystem of trust. To enable machines to give normative
reasons for their decisions and actions means to capacitate them of
communicative action \parencite{habermas81:_theor_handel}, or at least
to engage in constitutive communication of social systems
\parencite{luhmann84:_sozial_system}. The capacity to give and take
reasons is a crucial step towards fully autonomous normative (moral
and legal) agency. Moreover, our research, in the long run, paves way for interesting further studies and experiments on integrated neuro-symbolic AI architectures and on the emergence of patterns of self-reflection in intelligent autonomous machines.

%\hr{Move somewhere else and brieflz discuss?: Related work includes \textcite{DBLP:conf/aaai/LyuYLG19}.}

\paragraph{Acknowledgement:} We thank David Fuenmayor and the anonymous reviewers for their helpful
comments to this work.

\printbibliography

%\end{document}

\tikzset{block/.style={draw, thick, text width=2cm , minimum
    height=1.3cm, align=center}, line/.style={-latex} }
 
\tikzset{
  font={\fontsize{11pt}{12}\selectfont}}
  
\tikzset{
    testpic/.pic={
  \node[block, fill=Blue!40, text width=10cm, minimum width=18cm, minimum height=3cm] (m1) {
  \begin{minipage}[t][2cm]{10cm}
  \textbf{M1:} \\ \textbf{\MI} \\ \hrule \vskip3ex
  ethico-legal governance of intelligent autonomous agents
  \end{minipage}
  }; 
  \node[block, below left=.5cm and -5cm of m1, fill=Orange!40, text width=6cm, minimum height=5cm] (m3) {
  \begin{minipage}[t][4cm]{5.5cm}
  \textbf{M3:} \\ \textbf{\MIII} \\ \hrule \vskip3ex
  pluralistic normative reasoning  \\[2ex] 
  rule-based normative reasoning \\[2ex] 
  integrated and guided by M4
  \end{minipage}
  };
  \node[block, below=.5cm of m1, fill=OliveGreen!30, text width=7cm, minimum height=5cm] (m2) {
  \begin{minipage}[t][4cm]{6.5cm}
  \textbf{M2:} \\ \textbf{\MII} \\ \hrule \vskip3ex
  ethico-legal upper-level ontology \\[2ex] 
  value ontology (moral ``grammar'') \\[2ex] 
  ethico-legal regulation (code)  
  \end{minipage}
  };
  \node[block, below right=.5cm and -5cm of m1, fill=Orange!40, text width=6cm, minimum height=5cm] (m4) {
  \begin{minipage}[t][4cm]{5.5cm}
  \textbf{M4:} \\ \textbf{\MIV} \\ \hrule \vskip3ex
  ethico-legal reasoning net   \\[2ex] 
  interpretable AI to inform M3 
  \end{minipage}
  };
  \node[block, below=.5cm of m2, fill=Yellow!20, text width=10cm, minimum width=12cm, minimum height=4cm] (m5) {  
  \begin{minipage}[t][3cm]{10cm}
  \textbf{M5:} \\ \textbf{\MV} \\ \hrule \vskip3ex
  human-understandable rational arguments  \\[2ex] 
  human-centered interaction
  \end{minipage}
  };
  \node[block, below=.5cm of m5, fill=Blue!20, text width=10cm, minimum width=18cm, minimum height=3cm] (m6) {
  \begin{minipage}[t][2cm]{10cm}
  \textbf{M6:} \\ \textbf{\MVI} \\ \hrule \vskip3ex
  access {at scale} with little
  risk and minimal costs
  \end{minipage}
  };
  \node[block, above right=2cm and 1cm of m4, fill=black!20, text width=11cm, minimum width=13cm, minimum height=3cm, rotate=-90] (m7neu) {
  \begin{minipage}[t][2cm]{10cm}
  \textbf{M7:} \\ \textbf{\MVII} \\ \hrule \vskip3ex
  grand vision (top-down) \& module specific (bottom-up) 
  \end{minipage}
  }; 
 
 \begin{scope}[on background layer] 
 \node[draw, fill=black!20, inner xsep=20mm,inner
  ysep=5mm, fill opacity=0.3,fit=(m1)(m2)(m3)(m4)(m5)(m6)](all){};  
 \end{scope}  
  
 \begin{scope}[on background layer] 
 \node[draw, fill=Blue!40, inner xsep=5mm,inner
  ysep=-7mm, fill opacity=0.3,fit=(m1)(m2)(m3)(m4)(m5)(m6)](m1tom6){};
 \end{scope}
}}

\begin{figure}[htp!]
\resizebox{\textwidth}{!}{
    \rotatebox{90}{
        \begin{tikzpicture}
           \pic {testpic};
        \end{tikzpicture}
        }
    }   
\caption{Modular structure of \acronym\ research. \label{fig1}}
\end{figure}

\end{document}